\input{epsf}

\documentclass{article}
\usepackage{amssymb}
\usepackage{amsmath}
\usepackage{graphicx}
\usepackage{latexsym}

\newtheorem{ass}{Assumption}
\newtheorem{theorem}[ass]{Theorem}
\newtheorem{lemma}[ass]{Lemma}

\newcommand{\captionfonts}{\footnotesize}
\makeatletter  
\long\def\@makecaption#1#2{%
  \vskip\abovecaptionskip
  \sbox\@tempboxa{{\captionfonts #1: #2}}%
  \ifdim \wd\@tempboxa >\hsize
    {\captionfonts #1: #2\par}
  \else
    \hbox to\hsize{\hfil\box\@tempboxa\hfil}%
  \fi
  \vskip\belowcaptionskip}
\makeatother   

\begin{document}
 
\title{Results on the spectrum of R-Modes of slowly rotating relativistic 
stars}

\author{Horst R. Beyer \\
 Max Planck Institute for Gravitational Physics, \\
 Albert Einstein Institute, \\
 D-14476 Golm, Germany}

\date{\today}                                     

\maketitle

\begin{abstract}
The paper considers the spectrum of axial perturbations 
of slowly uniformly rotating general relativistic stars
in the framework of Y. Kojima. In a first step towards
a full analysis only the evolution equations are treated 
but not the constraint. Then it is found that the 
system is unstable due to a continuum of non real 
eigenvalues. In addition the resolvent of the associated generator 
of time evolution is found to have a {\it special structure} 
which was discussed in a previous paper. From this structure it 
follows the occurrence of a continuous part in the 
spectrum of oscillations at least if the system is 
restricted to a finite space as is done in most numerical 
investigations. Finally, it can be seen that higher order 
corrections in the rotation frequency can {\em qualitatively} 
influence the spectrum of the oscillations. As a consequence 
different descriptions of the star which are equivalent to 
first order could lead to different results with respect 
to the stability of the star.
\end{abstract}

\section{Introduction}
The discovery \cite{nanew,fm} of the instability of $r$-modes 
in rotating neutron stars by the emission of gravitational waves
via the Chandra\-sekhar-Friedman-Schutz 
(CFS) mechanism \cite{chandra,friedmanschutzb} found much 
interest among astrophysicists. That instability might be responsible  
for slowing down a rapidly rotating, newly-born neutron star 
to rotation rates comparable 
to the initial period of the Crab pulsar ($\sim$19 ms)
through the emission of current-quadrupole gravitational waves 
and would explain why only slowly-rotating pulsars are associated with 
supernova remnants \cite{anderssonkokkotasschutz,lindblomowenmorsink}. 
Also, while an initially rapidly rotating star spins down, 
an energy equivalent to roughly 1\% of a solar mass would be 
radiated in the form of gravitational waves, making the process 
an interesting source of detectable gravitational waves \cite{Owen98}. 
\newline
\linebreak
It was soon realized in a large number
of studies that many effects work against the growth of the r-mode 
like viscous damping, coupling to a crust, magnetic fields, 
differential rotation and exotic structure in the core of the neutron
star. Those effects could lead to a significant 
reduction of the impact of the instability or even to 
its complete suppression. For an account of those studies 
we refer to the recent reviews 
\cite{anderssonkokkotas,friedmanlockitch}.
\newline
\linebreak
Most of the results have been obtained using a newtonian description of 
the fluid and including 
radiation reaction effects by the standard 
multipole formula. Of course, such an ad hoc method
cannot substitute a fully general relativistic treatment of the system, 
but it was believed that it gives at least roughly correct results, both, 
qualitatively and quantitatively. However in a first step towards such 
a fully relativistic treatment it was shown in \cite{kojima} that the 
method misses important relativistic effects. Working in the low-frequency 
approximation Kojima could 
show that the frame dragging leads to the occurrence of a continuous part 
in the spectrum of the oscillations. This is qualitatively different 
from the newtonian case where this spectrum is `discrete'. \footnote{But note 
that depending on the equation of state still mode solutions can 
be found in the low-frequency approximation
\cite{lockitch, ruoffkokkotas, ruoff}.}
Mathematically, Kojima's arguments were not conclusive  
since drawn from an analogy to the equations 
occurring in the stability discussion of non-relativistic rotating 
ideal fluids \cite{chandra1,drazin} and because his mathematical reasoning 
still referred to `eigenvalues' (which generally don't exist in that case)  
rather than to values from a continuous part of a `spectrum'.
But soon afterwards in 
\cite{beyerkokkotas} K. Kokkotas and myself
provided a rigorous interpretation along with a  
proof of Kojima's claim. Indications for the continuous spectrum were also 
found in the subsequent numerical investigation \cite{ruoffkokkotas}. 
After that 
still the possibility remained that the result was an artefact of the used 
low-frequency approximation which in particular neglects gravitational 
radiation although numerical results in \cite{ruoff}
suggested that this is not the case. Also is 
Kojima's `master equation' (see (\ref{kojimasequation}))
time independent 
whereas mathematically it is preferable to start from 
a time dependent equation, 
because in this way it can be build on the known connection 
between the spectrum of the generator of time evolution 
and the stability of the system
\cite{hillephillips,kato,reedsimon}.
For these reasons we consider here 
Kojima's full equations for r-modes which include gravitational radiation
{\it but still neglect the coupling to the polar modes}. 
Indeed we will meet  
some surprises, related to the following. 
\newline
\linebreak
Due to lack of appropriate exact background solutions of 
Einstein's field equations the background model and its perturbation 
are expanded simultaneously into powers of the angular velocity  
$\Omega$ of the uniformly rotating star. In particular Kojima's equations are 
correct only to first order in $\Omega$. Of course, once the evolution 
equations of the perturbations are given, there is no room for neglecting 
any second nor higher order terms occuring in further computations.
The spectrum of the oscillations is determined by those equations and 
depending on that the system is stable or unstable. Now in the 
calculation it turns out that second order corrections in the coefficients 
of the evolution equations can {\em qualitatively} influence the 
spectrum of the 
oscillations. In particular 
continuous parts in the spectrum (both, stable and unstable) can come and 
go depending on such corrections. As a consequence different 
descriptions of the star which are equivalent to first order
could lead to different results with respect to its stability. 
Hence the decision on the 
stability would have to take into account 
such corrections which in 
turn would lead to considering a changed operator. In 
addition judging from the mathematical mechanism how this can happen it does 
not seem likely that this property of the equations is going to 
change in higher orders, which would ultimately question the 
expansion method as a proper means to investigate the stability of 
a rotating relativistic star. With respect to this point
further study is necessary, but still the results shed
some doubts on the appropriateness of the expansion method. 
   
\section{Mathematical Introduction}
Continuous spectra have been found in many cases 
in the past in the
study of differentially rotating fluids.
\cite{SV83,Balb84a}.
The continuous spectrum in these cases was again seen
for $r$-modes together with many interesting features
such as: the passage of low-order $r$-modes from the discrete part
into the continuous part as the differential rotation increases; and the
presence of low order discrete $p$-modes in the middle of the
continuous part in the more rapidly rotating disks \cite{SV83}.
\newline
\linebreak
{\em
The stars under consideration here have no differential rotation
and the existence of a continuous part of the spectrum is attributed
to the dragging of the inertial frames due to general relativity which is 
an effect not present in an newtonian description.}
\newline
\linebreak
Mathematically, the study of spectra containing continuous parts
requires a higher level of mathematical sophistication
than usual in astrophysics. Such parts can cause instabilities and they 
cannot be computed by straighforward mode calculations. Their occurrence 
makes it necessary to differentiate between `eigenvalues' (and the 
corresponding `modes') and the `spectrum' of the oscillations. The last 
depends on the introduction 
of a function space, a topology and the domain of definition of a linear 
operator (namely the generator of time evolution) analogous to quantum theory
and for this the use of subtle mathematics from functional analysis, 
operator theory and in particular `Semigroups of linear operators' 
is essential.
\newline
\linebreak
{\it Since the system considered here is dissipating (by gravitational radiation) the generator of time evolution has complex spectral values and 
is non self-adjoint. This complicates the investigation, because 
a general spectral theory for such operators comparable    
to that for self-adjoint operators is still far from 
existing. Indeed here there was not (even) found a `small' self-adjoint 
part of the generator which would have been suitable for 
applying the usual perturbation methods.} \footnote{
However note that such a method was successfully 
applied to Kojima's equation (\ref{kojimasequation})
for the low-frequency approximation 
\cite{beyerkokkotas}.}

\section{Kojima's Equations for R-modes}

Since the calculations here are based on the equations of Kojima
\cite{kojima0}
which are presented in detail there,
here we are going only briefly to describe the perturbation equations.
The star is assumed to be uniformly rotating with angular velocity
$\Omega\sim O(\epsilon)$ where
\begin{equation} 
\epsilon :=\Omega\sqrt{R^3/M}
\end{equation}
is small compared to unity. \footnote{
The assumption of slow rotation is considered to be a quite robust
approximation, because the expansion parameter
$\epsilon$ is usually very small and  the fastest
spinning known pulsar has  $\epsilon \sim 0.3$.} 
Here $M$ and $R$ are the mass and the 
radius of the star. Note that we use 
here and throughout the paper geometrical units 
$c=G=1$.
\newline
\linebreak
The background metric is given by:
\begin{equation}
g_{ab}^{(0)}=
-e^{\nu} dt^2 + e^{\lambda} d r^2 + r^2 (d\theta^2 + \sin^2 \theta
d \phi^2) 
-2\omega r^2 \sin^2 \theta d t d \phi \ ,
\end{equation}
where $\omega \sim O(\epsilon)$ describes the dragging of the inertial frame.
If we include the  effects of rotation only to order $\epsilon$ the
fluid is still spherical, because the deformation is of
order $\epsilon^2$  \cite{Hartle}.
The star is described by the standard Tolman-Oppenheimer-Volkov (TOV)
equations (cf Chapter 23.5 \cite{TOV}) plus an equation for $\omega$
\begin{equation}
\left(jr^2\varpi'\right)' -16\pi(\rho+p)e^\lambda j r^4 \varpi =0 \ ,
\end{equation}
where we have defined
\begin{equation}
\varpi = \Omega-\omega
\end{equation}
a prime denotes derivative with respect to $r$, and
\begin{equation} \label{j}
j=e^{-(\lambda+\nu)/2}.
\end{equation}
In the vacuum outside the star $\varpi$ can be written
\begin{equation}
\varpi=\Omega -{{2 J} \over r^3} \ ,
\end{equation}
where $J$ is the angular momentum of the star.
The function $\varpi$, both inside and outside the star is a function of $r$
only and continuity of $\varpi$ at the boundary (surface of the star, $r=R$)
requires that $\varpi_R'=6JR^{-4}$.
Additionally, $\varpi$ is monotonically increasing function of $r$
limited to
\begin{equation}
\varpi_0 \leq \varpi \leq \Omega ,
\end{equation}
where $\varpi_0$ is the value at the center.
\newline
\linebreak
The basic variables for describing r-modes propagating on the background
metric $g_{ab}^{(0)}$ are the functions $h_{0lm}(t,r),h_{1lm}(t,r)$
($r,\theta,\varphi$ spherical coordinates, $t$ time coordinate)
defined by expansion into spherical harmonics (imposing
the Regge-Wheeler gauge)
\begin{eqnarray}
h_{t \theta} &=& h_{\theta t} = -\sum_{l,m}h_{0lm}Y_{lm,\varphi}/sin\theta
\nonumber \\
h_{t \varphi} &=& h_{\varphi t} = \sum_{l,m}h_{0lm} sin\theta Y_{lm,\theta}
\nonumber \\
h_{r \theta} &=& h_{\theta r} = -\sum_{l,m}h_{1lm}Y_{lm,\varphi}/sin\theta
\nonumber \\
h_{r \varphi} &=& h_{\varphi r} = \sum_{l,m}h_{1lm} sin\theta Y_{lm,\theta}
\, \, , 
\end{eqnarray}
and the fluid perturbation $U_{lm}(t,r)$. Here
\begin{equation}
g_{ab} = g_{ab}^{(0)}+h_{ab} \, \, , 
\end{equation}
where $h_{ab}$ is the `small' perturbation.The dependance
of the basic variables on $l,m$ will be
suppressed in the following.
The evolution/constraint equation will be written in terms of
the following vector-valued variable
\begin{equation}
\vec{h} := 
\begin{pmatrix}
h_1 \\ h_0 \\ Z
\end{pmatrix}
\end{equation}
Then Kojima's equations for pure r-modes ({\it neglecting 
coupling to polar modes}) of a slowly 
and uniformly rotating 
general relativistic star take the following form 
($l \geq 2$):
\begin{equation} \label{evolh1} 
{\dot h_1} = - \, (D \vec{h})_{\, 1} := h_0' -Z 
\end{equation}

\begin{eqnarray} \label{evolh0a}
{\dot h_0} &=& - \, (D \vec{h})_{\, 2} \nonumber \\
           &:=& {e^{\nu}\over r^2}\left[2M+kr^3(p-\rho)\right]h_1
             + e^{\nu -\lambda} h_1' \nonumber \\
          && - {im\over \Lambda}
    \left[\Lambda \omega - 2 e^{-\lambda}r \omega'
      -\varpi\left( 2kr^2(\rho+p)+{4M\over r}-\Lambda \right)\right]h_0
      \\
 && + {im \over \Lambda}\varpi e^{-\lambda}r\left(rZ' -2h_0'\right)    
      -{im\over \Lambda}
     \left[ \omega'e^{-\lambda}r^2
        + \varpi\left( kr^3(\rho+p)-2re^{-\lambda}\right)\right] Z 
        \nonumber 
\end{eqnarray} 

\begin{eqnarray} \label{evolZa}   
{\dot Z}
      &=& -\, (D \vec{h})_{\, 3} \nonumber \\
    &:=& {e^{\nu} \over r^3}\left[r(\Lambda-2)+4M+2kr^3(p-\rho)\right] h_1
         +{2\over r}e^{\nu -\lambda} h_1'\nonumber \\
    && + 2{im\over \Lambda} r\varpi e^{-\lambda}Z'
     +{im \over \Lambda} \left(\Lambda \omega - 4e^{-\lambda}\varpi\right)h_0'
     \nonumber \\
     && +2{im\over \Lambda}\left[ \left(2e^{-\lambda}-kr^2(\rho+p)\right)\varpi
     -\Lambda \omega - e^{-\lambda} r \omega' \right]Z
    \nonumber \\
    &&+2{im\over \Lambda} \left[ (1+2e^{-\lambda})\omega' +
    \left(2kr^3(\rho+p)-\Lambda r +4 M \right){\varpi \over r^2} \right]h_0 
\, \, , 
\end{eqnarray}

where $\Lambda = l(l+1)$, $k = 4 \pi$, or  
in a more compact notation
\begin{equation} \label{evol}
\frac{d \vec{h}}{dt} = - D \vec{h} \, \, .
\end{equation}
In addition
we have:
\begin{eqnarray}
rZ' &-&2h_0'+\left[2-e^{\lambda}r^2 k(p+\rho) \right] Z -4 r e^{\lambda +\nu}U
\nonumber \\
     &+& {e^{\lambda}\over r^2}\left[4M-\Lambda r -2kr^3(p+\rho)\right]h_0
    -imr\omega h_1' \nonumber \\
&-& im\left[\left(2-kr^2e^\lambda(\rho+p)\right)\omega 
+  r\omega' {\Lambda +2
    \over \Lambda}\right]h_1 =0  \, \, .  
\label{constr}
\end{eqnarray} 
which gives the fluid velocity $U$ in terms of 
$h_0, h_1$ and $Z$. Since $U$ has to vanish outside
(\ref{constr}) {\it constrains the data} for (\ref{evol})
{\it outside the star}.

\subsection{Reminder on Results in the Low-Frequency 
Approximation}
Kojima \cite{kojima} investigates the r-modes 
of the system with low-frequency of the order $O(\epsilon)$.
He finds that the master equation governing those
oscillations is given by
\begin{equation} \label{kojimasequation}
 q \Phi +(\varpi-\mu)\left[v \Phi - \frac{1}{r^4j}
\left(r^4j \Phi ^{\prime} \right)^{\prime}\right] = 0 \ ,
\end{equation}
where
\begin{equation}
\Phi = \frac{h_0}{r^2} \ , 
\end{equation}
and
\begin{equation} \label{v}
v= \frac{e^\lambda}{r^2} (l-1)(l+2) \ , 
\end{equation}
\begin{equation} \label{q}
q= \frac{1}{r^4j} \left(r^4j \varpi^{\prime} \right)^{\prime}
  = 16\pi(\rho+p)e^\lambda \varpi \ ,  
\end{equation}
\begin{equation} \label{mu}
\mu = -\frac{l(l+1)}{2m}(\sigma-m\Omega).
\end{equation}
Mainly from its similarity 
with equations describing plane ideal newtonian fluids \cite{chandra1,drazin}
he concludes that the spectrum of the oscillations
is given by the singular values \footnote{i.e., zeros of 
the coefficient multiplying the highest order derivatives 
of the equation} of (\ref{kojimasequation})
inside the star,
i.e., by the range
\begin{equation}
\varpi_0 \leq \mu = - {{\ell(\ell+1)}\over {2m}}\left(\sigma -m\Omega\right)
\leq \varpi_{R} \, . 
\end{equation}
In fact in \cite{beyerkokkotas} it was proven that 
it is given by the larger set
\begin{equation} \label{spec0}
\varpi_0 \leq \mu = - {{\ell(\ell+1)}\over {2m}}\left(\sigma -m\Omega\right)
\leq \Omega \, .
\end{equation}
Note that such singular values are {\it not visible} in (\ref{evol}).

\section{The Evolution Equations}

In a first step we deal with the evolution equations only.
To formulate a well posed initial value problem
for the system (\ref{evol})
data will be taken from the 
Hilbert space \footnote{For the used notation   
compare the Conventions in the Appendix.}
\begin{equation}
X := L^2_{\Bbb{C}}(I,j) \times  L^2_{\Bbb{C}}(I,j) \times
 L^2_{\Bbb{C}}(I,r^2 j) \, \, , \, \, I := (0,\infty) \, \, ,
\end{equation}
where $j$ is defined by (\ref{j}).
Further we define an operator
$A:D(A) \rightarrow X$ by
\begin{equation}
D(A) := \left\{ \vec{h} \in  \left( 
C^1(I, \Bbb{C}) \times  C^2(I, \Bbb{C}) \times
 C^1(I, \Bbb{C})\right) \cap X: D \vec{h}\in X 
\right\}
\end{equation}
and
\begin{equation}
A \vec{h} :=  D \vec{h} \, \, , \, \, \vec{h} \in D(A) \, \, .
\end{equation}
Obviously, it can be seen  by partial integration that the adjoint operator
to $A$ is densely-defined \footnote{i.e, is defined on a subspace of 
$X$ which is in addition such that any 
element of $X$ is a limit point of that subspace.} and hence that
$A$ is closable, i.e., there is a unique `smallest' closed extension
of $A$ which is denoted by $\bar{A}$ in the following. 
In the following the system (\ref{evol}) is interpreted
as the abstract equation
\begin{equation} \label{evolabs}
\dot{\vec{h}}(t) = -\bar{A} \, \vec{h}(t) \, \, , \, \, t \in \Bbb{R} \, \, , 
\end{equation}
where the dot 
denotes the ordinary derivative of functions 
assuming values in $X$. The use of this formulation 
makes possible the application of the results in the field of   
`Semigroups of linear operators'.
In order that (\ref{evolabs}) has a unique solution for 
data from the domain of $\bar{A}$ it has to be proven 
that $\bar{A}$ is the generator of strongly continuous semigroup
(or group). That this is not just a technicality 
is already indicated 
by the fact that the system (\ref{evol}) can be seen to have
{\it complex characteristics} if the equation
\begin{equation} \label{restriction}
\frac{m^2}{\Lambda} \, \frac{r}{r-2M} \, \omega \varpi < \frac{1}{j^2} 
\end{equation} 
is violated. Hence in those cases it cannot be expected that 
$\bar{A}$ is such a generator. For this reason
(\ref{restriction}) is assumed to hold for now on. Note that
since 
\begin{equation}
\frac{1}{j^2} <  e^{-8 \pi \int_{0}^{R} r^{\prime}(p+\rho)e^{-\lambda} 
d r^{\prime}} 
\end{equation} 
and because of 
\begin{equation}
\frac{m^2}{\Lambda} \, \frac{r}{r-2M} \, \omega \varpi \sim
\frac{m^2}{\Lambda} \, \omega_{0}(\Omega - \omega_0) \quad  
\text{for $r \rightarrow 0$}
\end{equation}
and
\begin{equation}
\frac{m^2}{\Lambda} \, \frac{r}{r-2M} \, \omega \varpi \sim
\frac{2m^2 \Omega J}{\Lambda} \frac{1}{r^3} \quad
\text{for  $r \rightarrow \infty$}
\end{equation}
that there is a large range of values for  
the physical parameters where inequality (\ref{restriction})
is satisfied. But note also that 
we meet here for the first time a quantity of the order $O(\epsilon^2)$ 
which restricts the meaningfullness of (\ref{evol}) which itself is
correct only to the order $O(\epsilon)$.
\newline
\linebreak
If $\bar{A}$ is the generator of strongly continuous semigroup
(or group) --which is 
to expect-- its spectrum $\sigma(\bar{A})$ \footnote{given by all 
complex numbers $\sigma$ for which the corresponding 
map $\bar{A}- \sigma : D(\bar{A}) \rightarrow X$ is not \quad bijective}
tells us about the stability of the solutions of 
(\ref{evolabs}). For instance, if that spectrum contains 
values from the left half-plane of the complex plane then 
there are solutions of (\ref{evolabs}) which are exponentially 
growing. {\it Now the whole process of proving that $\bar{A}$
is such a generator would be greatly simplified by the usual
perturbation methods if $A$ could be split into a symmetric  
differential operator and into a `small' perturbation.
This was tried on a diagonalized form 
of the evolution equations (\ref{evol}), but unsuccessfully.
Indeed the problem there occurred in proving the `smallness' of the 
the perturbing operator. A further problem 
in that case was that assumptions of usually used theorems  
giving the asymptotics of solutions near $\infty$
which are needed for the construction of the
resolvent of the operator turned out to be not satisfied.
For these reasons that approach was not pursued any further.}
\newline
\linebreak
Usually, after the proof that an operator is 
the generator of a strongly continuous semigroup
the far more difficult problem occurs
in finding physically interesting properties
of its spectrum and here it might be thought
on first sight that this 
is analytically impossible, because of the complicated nature
of the evolution equations (\ref{evol}). But fortunately
in \cite{beyer3} there was found a whole class of operators 
\footnote{The operators of that class occurred as 
a natural generalization of the operators governing spheroidal 
oscillations of adiabatic spherical newtonian stars.}
which generally have a continuous part in their spectrum and where the
occurrence of that
part can be concluded from the {\it structure} of 
the resolvent. Indeed 
in the next section 
the {\it same structure} will be found in the resolvent 
of $\bar{A}$.     

\section{Construction of the Resolvent of the Generator} 
In a first step 
we try to invert the equations
\begin{equation} \label{spec}
(\bar{A} - \sigma ) \vec{h} = \vec{f}
\end{equation}
for $\vec{h} \in D(\bar{A})$ for any complex number $\sigma$ ($= i \sigma$ 
in Kojima's notation) 
and any  
continuous function  
\begin{equation}
\vec{f} =
\begin{pmatrix}
f_1 \\ f_2 \\ f_3
\end{pmatrix}
\end{equation}
assuming values in $\mathbb{C}^3$ and 
with {\it compact support} on the half line.
Any $\sigma$
for which there is some $\vec{f}$
such that (\ref{spec}) has not a unique solution 
is a spectral value. The unique inversion can fail for two reasons. Either 
there is a non trivial solution of the associated homogeneous equation
and hence $\sigma$ is an eigenvalue or otherwise 
there is no solution at all for some particular non trivial  
$\vec{f}$. Often the last happens not only for some 
isolated value of $\sigma$ but for a `continuous set' of values 
(like a real interval, a curve in $\mathbb{C}$ or even from an 
open subset of $\mathbb{C}$) leading to a `continuous part' in the 
spectrum. For all other values of $\sigma$ the inversion leads to a continuous 
(`bounded') linear operator on $X$.
\newline
\linebreak
Due to the special structure of $A$ 
(the orders of differentiation
inside $A$ vary in a special way) 
these equations can be decoupled
leading to a single second differential equation for 
$h_0$. This equation generalizes Kojima's equation (\ref{kojimasequation}). 
It is given by
\begin{eqnarray}
&&\left(p_1 - \frac{p_3 p_4}{q_4} \right) h_0^{\prime \prime} +
\left(p_2 - \left\{p_3 \left[ \left(\frac{p_4}{q_4}
\right)^{\prime} + \frac{q_3}{q_4} \right]  + \frac{q_2 p_4}{q_4} 
\right\}\right) 
h_0^{\prime} +  \nonumber \\  
&&\left( q_1 - \left[ p_3 \left(\frac{q_3}{q_4} \right)^{\prime}
+ \frac{q_2 q_3}{q_4}\right]\right)h_0 = g_3 := 
g_1 - \left[ p_3 \left(\frac{g_2}{q_4}\right)^{\prime} + \frac{q_2 g_2}{q_4} 
\right] \label{kojima}
\end{eqnarray}
where
\begin{eqnarray}
p_1 &=& \frac{im}{\Lambda} r^2 e^{-\lambda}\varpi \nonumber \\
p_2 &=& -\frac{im}{\Lambda}\left[ 
\omega^{\prime} e^{-\lambda} r^2 + kr^3 \varpi \left(p + \rho \right) \right] 
\nonumber \\
p_3 &=& e^{\nu -\lambda} + \frac{im \sigma}{\Lambda} \varpi r^2 e^{- \lambda} 
\nonumber \\
p_4 &=& \frac{r}{2} \left( \sigma - im \omega \right)
\end{eqnarray}

\begin{eqnarray}
q_1 &=& - \frac{im}{\Lambda} \left[ 
\Lambda \omega -2 e^{-\lambda}r \omega^{\prime}
- \varpi \left( 
2kr^2(p + \rho)-\frac{4M}{r}-\Lambda\right) 
+ \frac{i\Lambda \sigma}{m}\right] \nonumber \\
q_2 &=& \frac{e^{\nu}}{r^2} \left[ 2M + k r^3 (p-\rho)\right]
-  \frac{im \sigma}{\Lambda}
\left[  
\omega^{\prime}e^{-\lambda}r^2 + \varpi
\left(
k r^3 (p+\rho) - 2 r e^{-\lambda} \right)\right] \nonumber \\
q_3 &=& - \left[ 
\sigma - \frac{im}{\Lambda} \left(r\omega^{\prime} + 
\frac{8M}{r}\varpi + \Lambda \omega \right)\right] \nonumber \\
q_4 &=& \frac{e^{\nu}}{2r}(\Lambda-2) + 
\left(\frac{\sigma}{2}-im\omega \right) \sigma r 
\end{eqnarray}
and
\begin{eqnarray}
g_1 &=& - \frac{im}{\Lambda} r^2 e^{-\lambda} \varpi f_1^{\prime} + 
\frac{im}{\Lambda} \left[ 
\omega^{\prime} r^2 e^{-\lambda} +
\varpi \left( k r^3 (p+\rho)-2 r e^{-\lambda}\right)\right]f_1 -f_2  
\nonumber \\ 
g_2 &=&
-r \left( \frac{\sigma}{2}-im\omega\right)f_1 
+ f_2 - \frac{r}{2}f_3 \, \, .
\end{eqnarray}
From $h_0$ the functions $h_1$ and $Z$ can be computed by
\begin{eqnarray} \label{h1(h0),Z(h0,h1)}
h_1 &=& \frac{g_2}{q_4} - \frac{p_4}{q_4} h_0^{\prime} - \frac{q_3}{q_4} h_0 
\nonumber \\
Z &=& h_0^{\prime} + \sigma h_1 + f_1 \, \, .
\end{eqnarray}
Note that $q_4$ vanishes if only if 
\begin{equation} 
\sigma = i \left[ m \omega \pm \sqrt{m^2 \omega^2 + 
(\Lambda -2) \frac{e^{\nu}}{r^2}}\right]
\end{equation}
for some $r > 0$ and that 
\begin{equation}
p_1 q_4 - p_3 p_4 = - \frac{r}{2} e^{\nu - \lambda} 
\left[ \left( 1 - \frac{m^2}{\Lambda}\omega \varpi r^2 e^{-\nu}\right) \sigma
- im\left(\Omega - \frac{2}{\Lambda}\varpi \right)\right]
\end{equation}
vanishes if and only if
\begin{equation} \label{kojimaspec}
\sigma = \frac{im \left( \Omega - \frac{2}{\Lambda}\varpi \right)}
{1 -\frac{m^2}{\Lambda}\omega \varpi r^2 e^{-\nu}} 
\end{equation}
for some $r > 0$. 
Note that the last formula gives {\it up to 
first order in} $m$ exactly the values of the continuous spectrum found for 
(\ref{kojimasequation}). Also note that the denominator 
in (\ref{kojimaspec}) is greater than zero, because of  
\begin{equation}
 e^{\nu + \lambda} 
\left( 1 -\frac{m^2}{\Lambda}\omega \varpi r^2 e^{-\nu} \right) =
\frac{1}{j^2} - \frac{m^2}{\Lambda} \, \frac{r}{r-2M} \, \omega \varpi 
\end{equation}
and condition (\ref{restriction}) demanding that the 
right hand side of the last 
equation is greater than zero.\footnote{Remember that 
the last condition was imposed to exclude the occurrence 
of complex characteristics for the evolution equations (\ref{evol}).}
\newline
\linebreak
Hence both functions vanish only for purely imaginary $\sigma$.
So the equations are non singular for non purely imaginary $\sigma$
and this case is considered in the following.
We denote by $P_1$ the coefficient of
the leading order derivative in (\ref{kojima}) and by  $P_2$ the 
coefficient of the first order derivative
\begin{eqnarray}
P_1 &=& p_1 - \frac{p_3 p_4}{q_4} \, \, , \nonumber \\
P_2 &=& p_2 - \left\{p_3 \left[ \left(\frac{p_4}{q_4}
\right)^{\prime} + 
\frac{q_3}{q_4} \right]  + \frac{q_2 p_4}{q_4} \right\} \, \, .
\end{eqnarray}
If $h_{0l}$ and $h_{0r}$ (here, `\,$l$\,'
stands for `\,left\,' and `\,$r$\,' for `\,right\,')
are linear independent 
solutions of the {\it homogeneous}
equation associated with (\ref{kojima}) which 
are square integrable
near $0$ and near $\infty$, respectively, $h_0$
is given by 
\begin{eqnarray} \label{reph0}
h_0 (r) &=& 
\frac{1}{C} \left[ 
h_{0l}(r) \int_{r}^{\infty} 
h_{0r}(r^{\, \prime}) g_3(r^{\, \prime})
K (r^{\, \prime})
d r^{\, \prime}  \right. \nonumber \\
&& \qquad + \left.
h_{0r}(r) \int_{0}^{r} 
h_{0l}(r^{\, \prime})
g_3(r^{\, \prime}) K (r^{\, \prime})
d r^{\, \prime} 
\right] \, \, , 
\end{eqnarray} 
where $C$ is a non zero constant defined by
\begin{equation} \label{h0}
h_{0l}(r) h_{0r}^{\prime}(r) - h_{0r}(r) h_{0l}^{\prime}(r) =
C \, exp\left( - \int_{}^{r} \frac{P_2(r^{\, \prime})}{P_1(r^{\, \prime})} 
d r^{\, \prime}
\right)
\end{equation}
and 
\begin{equation} \label{K}
K(r) = \frac{1}{P_1(r)} \, 
\exp\left(\int_{}^{r} \frac{P_2(r^{\, \prime})}{P_1(r^{\, \prime})} 
d r^{\, \prime} \right) \, \, .
\end{equation}
The lower constant of integration has to be the same in the last two 
formulas. It is kept fixed in the following although its precise value
does not enter the formulas in any essential way. 
Note that the inhomogenity $g_3$ in (\ref{reph0})
allows the following decomposition into terms containing derivatives 
of the components of $\vec{f}$ and terms without such  
derivatives by
\begin{equation}  \label{g3}
g_3 = - \frac{im}{\Lambda}r^2 e^{-\lambda} \varpi f_1^{\, \prime} 
- p_3 \left(\frac{g_2}{q_4}
\right)^{\prime} + g_4 \, \, , 
\end{equation} 
where
\begin{equation}  
g_4 = \frac{im}{\Lambda} \left[ 
\omega^{\, \prime} r^2 e^{-\lambda} + 
\varpi \left( 
kr^3(p+\rho)-2re^{-\lambda}
\right)
\right]f_1
-f_2 - \frac{q_2}{q_4} g_2 \, \, .
\end{equation}
Finally from (\ref{h0}),(\ref{g3}) we get by 
partial integration
\begin{eqnarray} \label{h0fin}
h_0 (r) &=&   
\frac{im}{\Lambda} \frac{1}{C} \left[ 
h_{0l}(r) \int_{r}^{\infty} 
\left( h_{0r} r^2 e^{-\lambda} \varpi K \right)^{\, \prime}(r^{\, \prime}) 
f_1(r^{\, \prime})
d r^{\, \prime} + \right. \nonumber \\
&& \, \, \qquad \, \, \,  \left.
h_{0r}(r) \int_{0}^{r} 
\left( h_{0l} r^2 e^{-\lambda} \varpi K \right)^{\, \prime}(r^{\, \prime}) 
f_1(r^{\, \prime})
d r^{\, \prime} 
\right]   \nonumber \\
&& + \frac{1}{C} \left[ 
h_{0l}(r) \int_{r}^{\infty} 
h_{0r}^{\, \prime}(r^{\, \prime}) \frac{p_3}{q_4}(r^{\, \prime}) 
g_2(r^{\, \prime})
K (r^{\, \prime})
d r^{\, \prime} + \right. \nonumber \\
&& \qquad \,   \left. 
h_{0r}(r) \int_{r}^{\infty} 
h_{0l}^{\, \prime}(r^{\, \prime}) \frac{p_3}{q_4}(r^{\, \prime}) 
g_2(r^{\, \prime})
K (r^{\, \prime})
d r^{\, \prime} 
\right]   \nonumber \\ 
&& + \frac{1}{C} \left[ 
h_{0l}(r) \int_{r}^{\infty} 
h_{0r}(r^{\, \prime}) \frac{(p_3 K)^{\, \prime}}{q_4}(r^{\, \prime}) 
g_2(r^{\, \prime})
d r^{\, \prime}  + \right. \nonumber \\
&& \qquad \,   \left.
h_{0r}(r) \int_{r}^{\infty} 
h_{0l}(r^{\, \prime}) \frac{(p_3 K)^{\, \prime} }{q_4}(r^{\, \prime}) 
g_2(r^{\, \prime})
d r^{\, \prime} 
\right]  \nonumber \\ 
&& + \frac{1}{C} \left[ 
h_{0l}(r) \int_{r}^{\infty} 
h_{0r}(r^{\, \prime}) g_4(r^{\, \prime})
K (r^{\, \prime})
d r^{\, \prime} + \right. \nonumber \\
&& \qquad \,   \left. h_{0r}(r) \int_{0}^{r} 
h_{0l}(r^{\, \prime})
g_4(r^{\, \prime}) K (r^{\, \prime})
d r^{\, \prime} 
\right] 
\end{eqnarray}
Further it follows from 
(\ref{reph0}), (\ref{h0}) and (\ref{K}) 
that 
\begin{eqnarray} \label{reph0prime}
h_0^{\, \prime} (r) &=& 
\frac{1}{C} \left[ 
h_{0l}^{\, \prime}(r)\int_{r}^{\infty} 
h_{0r}(r^{\, \prime}) g_3(r^{\, \prime})
K (r^{\, \prime})
d r^{\, \prime} +  \right. \nonumber \\
&& \quad \, \, \left. h_{0r}^{\, \prime}(r) 
\int_{0}^{r}
h_{0l}(r^{\, \prime})
g_3(r^{\, \prime}) K (r^{\, \prime})
d r^{\, \prime} 
\right] 
\end{eqnarray} 
and hence by (\ref{h0}), (\ref{g3}) and partial integration
\begin{equation} \label{h0prime}
h_0^{\, \prime}(r) =  
h_{0I}^{\, \prime}(r) 
- \frac{p_3}{q_4 P_1} g_2 
- \frac{im}{\Lambda} \frac{r^2 e^{-\lambda} \varpi}{P_1} f_1 \, \, , 
\end{equation}
where
\begin{eqnarray} \label{h0Iprime}
h_{0I}^{\, \prime}(r) &:=&  
\frac{im}{\Lambda} \frac{1}{C} \left[ 
h_{0l}^{\, \prime}(r) \int_{r}^{\infty} 
\left( h_{0r} r^2 e^{-\lambda} \varpi K \right)^{\, \prime}(r^{\, \prime}) 
f_1(r^{\, \prime})
d r^{\, \prime} + \right. \nonumber \\
&& \qquad \, \, \, \left.
h_{0r}^{\, \prime}(r) \int_{0}^{r} 
\left( h_{0l} r^2 e^{-\lambda} \varpi K \right)^{\, \prime}(r^{\, \prime}) 
f_1(r^{\, \prime})
d r^{\, \prime} 
\right]   \nonumber \\
&& + \frac{1}{C} \left[ 
h_{0l}^{\, \prime}(r) \int_{r}^{\infty} 
h_{0r}^{\, \prime}(r^{\, \prime}) \frac{p_3}{q_4}(r^{\, \prime}) 
g_2(r^{\, \prime})
K (r^{\, \prime})
d r^{\, \prime} + \right. \nonumber \\
&& \qquad \left. h_{0r}^{\, \prime}(r) \int_{r}^{\infty} 
h_{0l}^{\, \prime}(r^{\, \prime}) \frac{p_3}{q_4}(r^{\, \prime}) 
g_2(r^{\, \prime})
K (r^{\, \prime})
d r^{\, \prime} 
\right]   \nonumber \\ 
&& + \frac{1}{C} \left[ 
h_{0l}^{\, \prime}(r) \int_{r}^{\infty} 
h_{0r}(r^{\, \prime}) \frac{(p_3 K)^{\, \prime}}{q_4}(r^{\, \prime}) 
g_2(r^{\, \prime})
d r^{\, \prime}  +  \right. \nonumber \\
&& \qquad \left. h_{0r}^{\, \prime}(r) \int_{r}^{\infty} 
h_{0l}(r^{\, \prime}) \frac{(p_3 K)^{\, \prime} }{q_4}(r^{\, \prime}) 
g_2(r^{\, \prime})
d r^{\, \prime} 
\right]  \nonumber \\ 
&& + \frac{1}{C} \left[ 
h_{0l}^{\, \prime}(r) \int_{r}^{\infty} 
h_{0r}(r^{\, \prime}) g_4(r^{\, \prime})
K (r^{\, \prime})
d r^{\, \prime} + \right. \nonumber \\
&& \qquad \left. h_{0r}^{\, \prime}(r) \int_{0}^{r} 
h_{0l}(r^{\, \prime})
g_4(r^{\, \prime}) K (r^{\, \prime})
d r^{\, \prime} 
\right] \, \, . 
\end{eqnarray}
Finally, using (\ref{h0prime}) in (\ref{h1(h0),Z(h0,h1)}) and some calculation
leads to 
\begin{equation} \label{h1}
h_1 = h_{1I} +
\frac{im}{\Lambda}\frac{r^2 e^{-\lambda}\varpi}{q_4 P_1} 
\left(
\frac{im\omega r}{2}f_1 + f_2 - \frac{r}{2} f_3 
\right)
\end{equation}
and
\begin{equation} \label{Z}
Z = h_{0I}^{\prime} + \sigma h_{1I} 
 -\frac{e^{\nu - \lambda}}{q_4 P_1} 
\left(
\frac{im\omega r}{2}f_1 + f_2 - \frac{r}{2} f_3 
\right)
\, \, ,
\end{equation}
where
\begin{equation} \label{h1I}
h_{1I} = - \frac{p_4}{q_4} h_{0I}^{\prime} - \frac{q_3}{q_4} h_0 \, \, .
\end{equation}
Note that the structure of these 
formulas for $h_1, Z$ is similar to that of formula (\ref{reph0})
for $h_0$, with one important difference. Apart from terms containing 
integrals both include an additive term which does not involve integration.
They are  
\begin{eqnarray}
&&\frac{im}{\Lambda}\frac{r^2 e^{-\lambda}\varpi}{q_4 P_1} 
\left(
\frac{im\omega r}{2}f_1 + f_2 - \frac{r}{2} f_3 
\right) \quad \text{for $h_1$} \nonumber \\
&& -\frac{e^{\nu - \lambda}}{q_4 P_1} 
\left(
\frac{im\omega r}{2}f_1 + f_2 - \frac{r}{2} f_3 
\right) \qquad \, \, \, \, \, \, \text{for $Z$}  \, \, .
\end{eqnarray}
Note that both factors multiplying the brackets
{\it diverge} at the values of $\sigma$ given by (\ref{kojimaspec}).
So here we recover again {\it up to first order in m} 
the values of the continuous spectrum found for 
(\ref{kojimasequation}). 
It is known from \cite{beyer3} that such values 
become part of the spectrum if the operator is
considered on a compact interval.
\footnote{See the proof of Theorem 17 in that paper.
Basis for this Lemma 2 in the Appendix and the compactness
of integral operators in (\ref{h0fin}) and  (\ref{h0Iprime}).} For the 
construction of the resolvent there are still needed 
solutions of the homogeneous equation corresponding to  
(\ref{kojima}) which are square integrable near $r=0$ and others
which are square integrable near $\infty$. The following Section 
investigates on their existence. 
 
\section{Asymptotics of the Homogeneous Solutions}

The result of the investigation
is as follows. In the following we drop the assumption that
$\sigma$ is non purely imaginary. 
There are \footnote{for e.g., according to the variant of Dunkel's theorem 
\cite{dunkel}
given in the Appendix.}
linearly independent solutions
$h_{01}, h_{02}$ satisfying
\begin{eqnarray} \label{specialsol1}
&&\lim_{r \rightarrow 0}  r^{-(l+1)}h_{01}(r) = 1  \cr
&&\lim_{r \rightarrow 0}  r^{-l}h_{01}^{\prime}(r) = l+1 \cr
&&\lim_{r \rightarrow 0}  r^{l}h_{02}(r) = 1  \cr
&&\lim_{r \rightarrow 0}  r^{l+1}h_{02}^{\prime}(r) = -l 
\end{eqnarray}
and for $\sigma$ different from 
\begin{equation} \label{exceptions}
0, \, \, \frac{im\Omega}{\Lambda^2}\Lambda(\Lambda-2), \, \,  
\frac{im\Omega}{\Lambda^2}\left[\Lambda(\Lambda-2) \pm 8mM\Omega\right]
\end{equation}
there are \footnote{for e.g., according to the proof of Theorem 1
in paragraph 8 of \cite{joergens} .}
linearly independent solutions 
\begin{eqnarray} \label{specialsol2}
h_{03}(r) &=&  r^{\rho_1}e^{\gamma_1 r }U_1(r) \nonumber \\
h_{04}(r) &=&  r^{\rho_2}e^{\gamma_2 r }U_2(r) \, \, ,
\end{eqnarray}
such that 
\begin{eqnarray}
&&\lim_{r \rightarrow \infty} U_n(r) = 1 \nonumber \\
&&\lim_{r \rightarrow \infty} U^{\, \prime}_n(r) = 0 \, \, .
\end{eqnarray}
for $n=1,2$.
Here $\gamma_1$, $\gamma_2$ are solutions of 
\begin{equation} \label{equation}
\gamma^2 -\frac{16m^2 M \Omega^2}{\Lambda^2} 
\frac{\sigma}{\sigma - \frac{im}{\Lambda}(\Lambda -2)\Omega } \gamma - \sigma^2
= 0 
\end{equation}
such that $\, \text{Re}\,(\gamma_1) \leq \, \text{Re}\,(\gamma_2)$, 
\begin{equation}
\rho_n = - \frac{\sigma_1 \gamma_n + \tau_1}{2\gamma_n + \sigma_0} \, \, ,
\end{equation}
for $n=1,2$, where
\begin{eqnarray}
&&\sigma_1 = \left\{32 m^2 M \Omega^2 \sigma \left[
- \left( \Lambda M + m^2 \Omega J\right)\sigma + imM \Omega(\Lambda -2)
\right] \right. \nonumber \\
&& \left. \, \, \, \,  \qquad - 2\Lambda \left[\Lambda \sigma - im (\Lambda -2)\Omega \right]^2
\right\} \left\{\Lambda^3 \left[ \sigma - \frac{im}{\Lambda}(\Lambda-2)\Omega\right]^2\right\}^{-1}
\quad ,
\nonumber \\
&& \tau_1 = -\frac{2 \sigma^2 \left[ 
(J \Omega m^2 +2 M \Lambda) \sigma -2iM \Omega \Lambda m  
\right]}
{\Lambda \left[ \sigma - \frac{im}{\Lambda}(\Lambda-2)\Omega\right]} \quad , 
\nonumber \\
&&\sigma_0 = -\frac{16m^2 M \Omega^2}{\Lambda^2} 
\frac{\sigma}{\sigma - \frac{im}{\Lambda}(\Lambda -2)\Omega } 
\quad . 
\end{eqnarray} 
Note that the presence of a second order 
term in (\ref{equation}) diverging near 
\begin{equation}
\sigma = \frac{im}{\Lambda}(\Lambda -2)\Omega \, \, ,
\end{equation} 
which is the newtonian frequency for r-modes (to first order 
in $\Omega$) as seen from an inertial observer \cite{beyerkokkotas}.
Despite of being second order that term becomes 
arbitrarily large near this frequency. 
\newline
\linebreak
Further, note that (\ref{h1(h0),Z(h0,h1)}), (\ref{specialsol1}), 
(\ref{specialsol2}) imply that the corresponding 
$h_{11},h_{12},h_{13},h_{14},Z_{1},$
$Z_{2},Z_{3},Z_{4}$ satisfy 
\begin{eqnarray} 
&&\lim_{r \rightarrow 0}  r^{-(l+2)}h_{11}(r) = 
- (\sigma - im \omega(0))(j(0))^2 / (l+2)
  \cr
&&\lim_{r \rightarrow 0}  r^{l-1}h_{12}(r) = 
 (\sigma - im \omega(0))(j(0))^2 / (l-1)
  \cr
&&\lim_{r \rightarrow 0}  r^{-l}Z_1(r) = l+1 \cr
&&\lim_{r \rightarrow 0}  r^{l+1}Z_2(r) = -l    
\end{eqnarray}
and 
\begin{eqnarray} 
&&\lim_{r \rightarrow \infty}  r^{-\rho{_1}}e^{-\gamma{_1}r}
h_{\, 13}(r) = - \gamma_{1} / \sigma \cr
&&\lim_{r \rightarrow \infty}  r^{-\rho{_2}}e^{-\gamma{_2}r}
h_{\, 14}(r) = - \gamma_{2} / \sigma \cr
&&\lim_{r \rightarrow \infty}  r^{1-\rho{_1}}e^{-\gamma{_1}r}
Z_{\, 3}(r) = 2 \cr
&&\lim_{r \rightarrow \infty}  r^{1-\rho{_2}}e^{-\gamma{_2}r}
Z_{\, 4}(r) = 2 \, \, \, \, .   
\end{eqnarray}
\newline
\linebreak
The outcome for the solutions near $0$ is, of course, 
exactly as expected. As a consequence for any 
$\sigma$ the triple
\begin{equation} 
\vec{h}_1 := {}^{t}(h_{11},h_{01},Z_{1})
\end{equation}
is in $X$ near $0$.
\newline
\linebreak
The analysis of the solutions of (\ref{equation}) 
uses Theorem (39,1) in \cite{marden} which generalizes 
the Routh-Hurwitz theorem to polynomials with
{\it complex} coefficients. For this 
solutions  
with $Re(\gamma) < 0 $, $Re(\gamma)> 0$, will be referred to as 
`stable' and `unstable', respectively. 
For the analysis define the discrimants
$\triangle_1$ and $\triangle_2$ of (\ref{equation}) by 
\begin{eqnarray}
\triangle_1 &:=& \frac{16m^2 M \Omega^2}{\Lambda^2} 
\left|\sigma - \frac{im}{\Lambda}(\Lambda - 2)\Omega \right|^{-2} 
\nonumber \\
&& \left\{ \sigma_1^2 + \left[ \sigma_2 - \frac{m}{2 \Lambda}
(\Lambda -2) \Omega \right]^2 - \frac{m^2}{4 \Lambda^2}(\Lambda -2)^2 
\Omega^2 \right\}
\end{eqnarray}
and
\begin{eqnarray}
&&\triangle_2 :=
- \left\{  
\frac{16m^2 M \Omega^2}{\Lambda^2} 
\left|\sigma - \frac{im}{\Lambda}(\Lambda - 2)\Omega \right|^{-2} 
(\sigma_1^2 + \sigma_2^2) \cdot
\right.\nonumber \\
&& \, \, \qquad \qquad \left.
 \left[ 
\sigma_1^2 + \sigma_2 \left(\frac{m}{\Lambda}(\Lambda-2)\Omega- \sigma_2
\right)
\right] \triangle_1 + 4 \sigma_1^2 \sigma_2^2
\right\} \, \, . 
\end{eqnarray}
Then if, both, $\triangle_1 \neq 0$ and $\triangle_2 \neq 0$ 
the number of zeros of  (\ref{equation}) in the open left half-plane
of the complex half-plane is given by the sign changes in 
the sequence $S=1,\triangle_1,\triangle_2$ and the number
of zeros in the open right half-plane is given by the number
of permanences of sign in this sequence.
Note that $\triangle_1$ is $>0$ ($<0$) outside (inside)
the circle
\begin{equation}
\sigma_1^2 + \left[ \sigma_2  
- \frac{m}{2\Lambda}(\Lambda -2)\Omega\right]^2
= \frac{m^2}{4\Lambda^2}(\Lambda - 2)^2 \Omega^2  \, \, .
\end{equation} 
So we consider the following cases.
\begin{itemize}
\item 
The first case (corresponding to regions I in Fig.1) 
considers the values of $\sigma$
in the complement 
of the closed strip $\mathbb{R} \times[0, m(\Lambda-2)\Omega / \lambda]$.
Here we have $\triangle_1 > 0$ and hence maximal one sign change 
in the sequence. In particular, there is no
change in sign near the imaginary axis. So 
there is maximally one stable solution 
and especially {\it no} stable solution near the imaginary 
axis.
\item
The second case (corresponding to region II in Fig 1) considers the
values of $\sigma$ in the  
open strip $\mathbb{R}\times \left(0, m(\Lambda-2)\Omega / \lambda \right)$. 
Here we have three subcases a), b) and c)
(corresponding to regions A, B and C, respectively, in Fig 1.)

Subcase a) considers  
those values with $\triangle_1 > 0$.
Obviously, this implies $\triangle_2 <  0$. As a consequence 
there is exactly one sign change in the sequence and
hence there are exactly one stable and one unstable 
solution.

Subcase b) considers those values with $\triangle_2 > 0$.
This implies  $\triangle_1 < 0$ and two sign changes in the 
sequence. Hence in this case there are only stable solutions.

The last subcase c) considers those values satisfying, both,
$\triangle_1 < 0$ and $\triangle_2 < 0$. Then there is only
one sign change and hence exactly one stable and one unstable 
solution.
\end{itemize}
Finally, by an application of Rouche's theorem (see e.g. Theorem 3.8
in Chapter V of \cite{conway})
follows that   
\begin{itemize} 
\item For 
\begin{equation} \label{estimate}
|\, \text{Re}(\sigma)| > \frac{|m|\,|\Omega|}{\Lambda}
\left( \Lambda -2 + \frac{8|m|M|\Omega|}{\Lambda} 
\right)
\end{equation}
there exist, both, a solution with $\gamma_1 < 0$ and with 
$\gamma_1 > 0$ and hence a stable and an unstable 
solution.
\end{itemize}
\begin{figure}
\centerline{\includegraphics[width=0.62\textwidth]{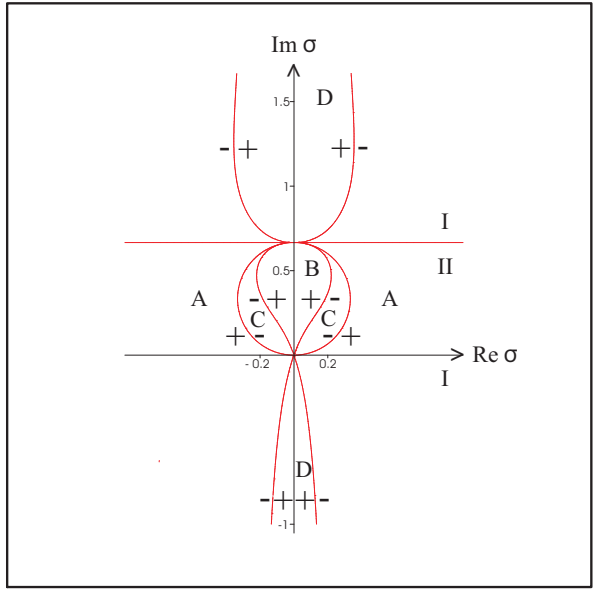}}
\caption{Sign changes in
S for $m=1,M=1,\Omega=1,\Lambda=6$.}
\label{fig1}
\end{figure}
The asymptotics near $\infty$
of  $h_{03}, h_{04}$ for $\sigma$ near 
and on the imaginary axis
is surprising. Expected was that in the complement
of imaginary axis there are always, both, a stable and an unstable solution
of (\ref{equation}) and on the imaginary axis 
that both solutions are purely 
imaginary. Indeed this would have been the case if 
the second order term in this equation were absent.
There one has either exponential growth of
both solutions (in Regions D) or exponential decay of both solutions
(Region B). As a consequence each $\sigma \in B$ -- different from the 
values given in (\ref{exceptions}) -- is an {\it eigenvalue} of 
$A$. Hence there is a {\it continuum of eigenvalues}
for $A$ in the open left half-plane and hence the 
evolution given by (\ref{evolh1}), (\ref{evolh0a})
and (\ref{evolZa}) is {\it unstable}.

\section{Discussion and Open Problems}
In the previous Section it was found that the spectrum of $\bar{A}$
contains a {\it continuum of unstable eigenvalues} leading to 
an unstable evolution given by (\ref{evolh1}), (\ref{evolh0a})
and (\ref{evolZa}). In addition there was found a continuum of 
values $\sigma$ (from Regions D in Fig.1
-- which include the spectral values found in the low-frequency approximation 
as well as (\ref{kojimaspec}) -- for which any
non trivial solution of the homogeneous equation corresponding to 
(\ref{spec}) growths
exponentially near $\infty$. Hence it is to expect that those 
values are also part of the spectrum of $\bar{A}$ and hence contribute
to the instability. Note in particular that 
their associated growth times displayed in Fig.1 can be of 
larger size than those for the found unstable 
eigenvalues (Region B in Fig.1) which suggests that the 
influence on the evolution dominantes over the influence 
of those eigenvalues.
\newline
\linebreak 
A further question is 
how the constraint (\ref{constr}) -- which has not been treated so far --
is likely to influence the spectrum. The constraint should lead to 
closed subspace of $X$ being invariant under $\bar{A}$. Then the spectrum 
of the restriction of $\bar{A}$ to that subspace is a subset of the 
spectrum of $\bar{A}$. This should
have the effect that the eigenvalues in region
$B$ (see Fig.1) are becoming discrete instead whereas the continuous 
part of the spectrum in Region $D$ might be left unchanged, because 
in general continuous spectra are less sensitive to such a operation.
In any case the singular structure of the resolvent at the values  
given by (\ref{kojimaspec}) will be unchanged having the effect that 
those values remain part of the spectrum at least when the 
system is restricted to finite space as is necessarily done in 
most numerical investigations. To decide these 
questions further study is needed. Note in this 
connection that the occurrence of an unbounded 
spectrum for $\bar{A}$ would not be a surprise, because 
for e.g., it is known that a self-adjoint operator has a bounded
spectrum if and only if it is defined on the whole 
Hilbert space and in general differential operators cannot 
be defined on the whole of a weighted $L^2$-space. Also would 
the occurrence of an infinitely extended continuous part in the spectrum 
of oscillations be not surprising since that is quite common for infinitely 
extended physical systems.   
\newline
\linebreak 
From a physical point of view the main worrying
feature of the results is their qualitative 
dependance on
{\it second order terms} like that in (\ref{equation}).
It is to expect that changes in second order to the coefficents 
in (\ref{evolh1}), (\ref{evolh0a}) and (\ref{evolZa})
influence that term which can lead to a qualitative change of the 
spectrum. As a consequence 
different descriptions of the star which are equivalent to first order
could lead to different statements 
on the stability of the star.
Hence to decide that stability such corrections of the coefficients
would have to be taken into account which in
turn would lead to considering a changed operator. In 
addition judging from the mathematical mechanism how the 
coefficients of the evolution equations influence the spectrum
-- namely through the asymptotics of the homogeneous solutions of 
(\ref{spec}) near $\infty$ --  
it might be suspected
that this property of the equations does not change in 
higher orders, which would ultimately question the 
expansion method as a proper means to investigate the stability of 
a rotating relativistic star. Also with respect to these points
further study is necessary, but still the results raise first
doubts whether the slow rotation approximation 
is appropriate for this purpose.  
\newline
\linebreak
A final interesting question to ask is whether a numerical 
investigation of the evolution given by (\ref{evolh1}), (\ref{evolh0a}) 
and (\ref{evolZa}) is capable of detecting the computed instabilities. 
This seems unlikely, because in that process space has to be 
`cutoff' near the singular points $r=0, \infty$  
and suitable local boundary conditions have to be posed at that 
the endpoints. But such a system is 
qualitatively different from the infinite system since for instance 
there the asymptotics of the homogeneous solutions (\ref{spec}) 
near $\infty$ does not play a role. Note that -- independent from the
used boundary conditions -- for such a
system the continuum of values given by the restriction of
(\ref{kojimaspec}) to the chosen interval
{\it are part of the spectrum} \footnote{
This is an easy consequence of Lemma 2
in the Appendix along with the compactness
of the integral operators involved in the representation of the resolvent
given by  (\ref{h0fin}) - (\ref{h1I}). Compare also the proof of
Theorem 17 in \cite{beyer3}.}
of the corresponding operator
as also is also indicated by the numerical investigation \cite{ruoff}.

\subsection*{Acknowledgements}
I would like to thank Kostas Kokkotas and Johannes Ruoff for many important 
discussions. Also I would like to thank B. Schmidt for reading the paper
and helpful suggestions.

\section{Appendix}
\subsection{Conventions}
The symbols ${\Bbb{N}}$, ${\Bbb{R}}$, ${\Bbb{C}}$ denote the 
natural numbers 
(including zero), all real numbers and all
complex numbers, respectively. \newline
\linebreak
To ease understanding we follow common abuse of notation and don't 
differentiate between coordinate maps and coordinates. For instance, 
interchangeably $r$ will denote some real number greater
than $0$ or the coordinate projection onto the open interval
$I := (0, \infty)$. The definition used
will be clear from the context. In addition we assume composition 
of maps (which includes addition, multiplication etc.) always to be   
maximally defined. So for instance the addition of two maps (if at all
possible) is defined on the intersection of the corresponding domains.
\newline
\linebreak
For each $k \in {\Bbb{N}} \setminus \{0\}$, $n \in {\Bbb{N}} \setminus \{0\}$ and 
each non-trivial open subset 
$M$ of ${\Bbb{R}}^n$ the symbol
$C^k(M,{\Bbb{C}})$ denotes the linear space of $k$-times 
continuously differentiable complex-valued functions on $M$. 
Further $C^k_{0}(M,{\Bbb{C}})$ denotes the subspace of
$C^k(M,{\Bbb{C}})$ consisting of those functions which in addition
have a compact support in $\Omega$.
\newline
\linebreak
Throughout the paper Lebesgue integration theory is used
in the formulation of \cite{riesznagy}.  
Compare also Chapter III in 
\cite{hirzebruchscharlau} and Appendix A in \cite{weidmann}.
To improve readability we follow common usage 
and don't differentiate between an 
almost everywhere (with respect to the chosen measure) defined 
function $f$ and the associated equivalence class
(consisting of all almost everywhere defined functions which differ 
from $f$ only on a set of measure zero). 
In this sense
$L_{C}^2\left(M,\rho\right)$, where $\rho$ is some strictly 
positive real-valued continuous function on $M$,     
denotes the Hilbert space of complex-valued, 
square integrable (with respect to the measure $\rho \, d^nx$) 
functions
on $M$. 
The scalar product 
$<| >$ on $L_{C}^2\left(M,\rho \right)$  is defined by
\begin{equation} \label{scalarproduct}
<f|g> := \int_{M}
   f^*
   g \, \rho \,  d^nx  \, \, ,
\end{equation}
for all $f,g \in L_{C}^2\left(M,\rho \right)$, where $^*$ denotes complex 
conjugation on ${\Bbb{C}}$. It is a standard result of functional analysis
that  $C^k_{0}(M,{\Bbb{C}})$ is dense in $L_{C}^2\left(M,\rho\right)$.
\newline
\linebreak
Finally, throughout the paper standard results and nomenclature of 
operator theory is used. For this compare standard 
textbooks on Functional analysis, e.g.,\cite{reedsimon} Vol. I,
\cite{riesznagy,yosida}.

\subsection{Auxiliary Theorems}

The variant of the theorem of Dunkel \cite{dunkel} 
(compare also \cite{Levinson,Bellman,Hille})
used in Section 4 is the following. 
\begin{theorem}: Let $n \in N$; $a, t_0 \in R$ with $a < t_0$; $\mu \in N$;
$\alpha_{\mu} := 1$ for  $\mu=0$ and
$\alpha_{\mu} := \mu$ for  $\mu \neq 0$. 
In addition let $A_0$ be a diagonalizable  complex  $n \times n$ 
matrix and $e^{\prime}_{1} , \dots,   e^{\prime}_{n}$
be a basis of $C^{n}$ consisting of eigenvectors of  $A_0$. 
Further, for each $j \in \{1,\cdots,n\}$ let $\lambda_{j}$ 
be the eigenvalue corresponding to  $e^{\prime}_{j}$ and $P_{j}$ 
be the matrix representing the projection of $C^{n}$
onto $C.e^{\prime}_{j}$ with respect to the canonical basis of $C^{n}$.
Finally, let $A_{1}$ be a continuous map from $(a, t_0)$ into the complex 
$n \times n$ matrices $M(n \times n,C)$ for which there is 
a number $c \in (a,t_0)$ such that the restriction 
of $A_{1jk}$ to $[c,t_0)$ is Lebesgue integrable for each 
$j,k \in {1,..., n}$. 
\newline
\newline
Then there is a $C^{1}$ map $R:(a,t_{0}) \rightarrow M(n \times n,C)$
with $lim_{t \rightarrow 0}R_{jk}(t) = 0$ for each $j,k \in {1,\dots,n}$ 
and such that $u:(a,t_{0}) \rightarrow M(n \times n,C)$ defined by
\begin{eqnarray} \label{u}
u(t):=
\left\{
 \begin{array}{ll}
  \sum^{n}_{j=1}(t_0-t)^{-\lambda_{j}}\cdot (E+R(t))\cdot P_{j} 
  \ \mbox{for $\mu=0$} \cr
  \sum^{n}_{j=1}exp(\lambda_{j}(t_0-t)^{-\mu})\cdot (E+R(t))\cdot P_{j} 
   \ \mbox{for $\mu \neq 0$ }
 \end{array} \right. \,
\end{eqnarray}
for all $t \in (a,t_{0}) $
(where $E$ is the $n \times n $ unit matrix),  maps into the 
invertible  $n \times n $ matrices and satisfies
\begin{equation} \label{asymptoticofu}
u^{\prime}(t) = \left( \frac{\alpha_{\mu}}{(t_0-t)^{\mu+1}}A_{0} + A_{1}(t) 
\right) \cdot u(t)
\end{equation}
for each $t \in (a,t_{0})$.
\end{theorem}

\begin{lemma}
Let $X$ be a non-trivial Hilbert space with scalar product
$<\,|\,>$ and induced norm $\|\,\|$ and let be $A : D(A) \rightarrow X $
a densely-defined, linear and closable operator in $X$. Further
let be $\sigma \in {\Bbb{C}}$ such that $A-\sigma$ is injective. Finally,
let be $\mu \in {\Bbb{C}^{*}}$ and let be 
$\eta_0, \eta_1,\dots$ a sequence of 
elements of $Ran(A-\sigma)$ for which there is 
an $\varepsilon \in (0,\infty)$ 
such that 
$\|\eta_{\nu}\| \geq
\varepsilon$  
for every $\nu \in {\Bbb{N}}$ and
\begin{equation} \label{limit}
\lim_{\nu \rightarrow \infty} \left[ 
\left( A - \sigma \right)^{-1}\eta_{\nu} -\mu \eta_{\nu} 
\right] = 0 \, \, .
\end{equation}
Then $\sigma + \mu^{-1}$ is in the spectrum of the 
closure $\bar{A}$ of $A$.
\end{lemma}
{\bf Proof:} The proof is indirect. Assume otherwise that 
$\sigma + \mu^{-1}$ is not part of the spectrum of $\bar{A}$.
Then $\bar{A}-(\sigma + \mu^{-1})$ is in particular bijective
with a bounded inverse $\left(\bar{A}-(\sigma + \mu^{-1})\right)^{-1}$.
Then we have for every $\nu \in {\Bbb{N}}$
\begin{equation}
\left(\bar{A}-
(\sigma + \mu^{-1})\right)\left( A - \sigma \right)^{-1}\eta_{\nu} =
\eta_{\nu} - \mu^{-1}\left( A - \sigma \right)^{-1}\eta_{\nu}
\end{equation} 
and hence also
\begin{eqnarray}
\left(\bar{A}-
(\sigma + \mu^{-1})\right)^{-1}\left( A - \sigma \right)^{-1}\eta_{\nu} &=&
\mu \left(\bar{A}-(\sigma + \mu^{-1})\right)^{-1} \eta_{\nu} - \nonumber \\
&& \mu \left(\bar{A}- \sigma \right)^{-1} \eta_{\nu} \, \, .
\end{eqnarray}
Using this it follows from (\ref{limit}) and the continuity of  
$\left(\bar{A}-(\sigma + \mu^{-1})\right)^{-1}$ that
\begin{eqnarray}
&& \lim_{\nu \rightarrow \infty} 
\left\{ 
\left(\bar{A}-
(\sigma + \mu^{-1})\right)^{-1}
\left( A - \sigma \right)^{-1}\eta_{\nu} 
-\mu \left(\bar{A}-
(\sigma + \mu^{-1})\right)^{-1}\eta_{\nu}
\right\} = \nonumber \\
&& - \mu \lim_{\nu \rightarrow \infty}  
\left( A - \sigma \right)^{-1}\eta_{\nu}  
 = 0
\end{eqnarray} 
and hence by (\ref{limit}) that 
\begin{equation}
\lim_{\nu \rightarrow \infty} \eta_{\nu} = 0 \, \, . 
\end{equation}
The last is in contradiction to the assumption there is 
an $\varepsilon \in (0,\infty)$ such that 
$\|\eta_{\nu}\| \geq
\varepsilon$ 
for all $\nu \in {\Bbb{N}}$.
Hence the Lemma is proven.$_\Box$


\begin{thebibliography}{99}

\bibitem{nanew} Andersson N 1998,  
{\em A New Class of Unstable Modes of Rotating Relativistic Stars},
ApJ, 502, 708-713.

\bibitem{anderssonkokkotas} Andersson N, Kokkotas K D 2001
{\em The r-mode instability in rotating neutron stars},
Int. J. Mod. Phys. D10, 381-442,
gr-qc/0010102.

\bibitem{anderssonkokkotasschutz} Andersson N, Kokkotas K D and
Schutz B F 1999 {\em Gravitational radiation limit on the spin of 
young neutron stars}, ApJ, 510, 846-853. 

\bibitem{AKS98b}
        Andersson N, Kokkotas K D and Stergioulas N 1999 
        {\em On the relevance of the r-mode instability for 
        accreting neutron stars and white Dwarfs},
        ApJ, 516, 307-314, gr-qc/0109065.
 

\bibitem{Balb84a}
        Balbinski E 1984, 
{\em The continuous spectrum in differentially rotating perfect fluids - A model with an analytic solution},
M.N.R.A.S., 209, 145-157.

\bibitem{Bellman} 
      Bellman R., 1949, 
       {\em  A survey of the theory of the boundedness, 
            stability, and asymptotic behaviour of solutions of linear 
            and nonlinear differential and difference equations} 
           (Washington DC: NAVEXOS P-596, Office of Naval Research).

\bibitem{beyer1} Beyer H R 1995 {\em The spectrum of radial 
adiabatic stellar oscillations} J. Math. Phys., 36, 4815-4825.

\bibitem{beyer2} Beyer H R 1995 {\em The spectrum of  
adiabatic stellar oscillations} J. Math. Phys., 36, 4792-4814.

\bibitem{beyer3} Beyer H R 2000 `On some vector analogues
of Sturm-Liouville operators' in: {\em Mathematical
analysis and applications}, T. M. Rassias (ed.), (Palm Harbor: Hadronic Press),
11-35.  

\bibitem{beyerkokkotas} Beyer H R and Kokkotas K D 1999 
{\em On the r-mode spectrum of relativistic stars} 
Mon. Not. R. Astron. Soc., 308, 745-750.

\bibitem{beyerschmidt} Beyer H R and Schmidt B G 1995 
{\em Newtonian stellar oscillations} Astron. Astrophys., 296, 722-726.

\bibitem{chandra1} Chandrasekhar S 1981 {\em Hydrodynamic and
Hydromagnetic Stability} (New York: Dover).

\bibitem{chandra} Chandrasekhar S 1970, 
{\em Solutions of Two Problems in the Theory of Gravitational Radiation}
Phys. Rev. Lett., 24, 611-615.

\bibitem{conway} Conway J B 1995 {\em Functions of a complex variable I}
2nd ed. (New York: Springer).

\bibitem{drazin}
Drazin P G and Reid W H 1981 {\em Hydrodynamic stability}
Cambridge: Cambridge University Press).

\bibitem{dunkel} 
        Dunkel O., 1912, 
        Am. Acad. Arts Sci.Proc., 38, 341. 

\bibitem{friedmanlockitch} Friedman J L and Lockitch K H 2001,
{\em Implications of the r-mode instability of rotating relativistic stars},
Review to appear in the proceedings of the 9th Marcel Grossman Meeting, 
World Scientific, ed. V. Gurzadyan, R. Jantzen, R. Ruffini, gr-qc/0102114.

\bibitem{friedmanschutzb} Friedman J L and Schutz B F 1978
{\em Secular instability of rotating newtonian stars}
ApJ, 222, 281-296.

\bibitem{fm}
        Friedman J L and Morsink S
         1998, {\em Axial Instability of Rotating Relativistic Stars},
         ApJ, 502, 714-720.

\bibitem{Hartle} 
        Hartle J B 1967, {\em Slowly Rotating Relativistic Stars: I. Equations 
        of Structure},
        ApJ, 150, 1005-1029.

\bibitem{Hille}  
       Hille E., 1969, 
      {\em Lectures on ordinary differential equations} 
       (Reading: Addison-Wesley). 

\bibitem{hillephillips} Hille E and Phillips R S 1957 {\em Functional
 Analysis and Semi-Groups} (Providence: AMS).

\bibitem{hirzebruchscharlau} 
       Hirzebruch F and Scharlau W 1971  
       {\em  Einf\"{u}hrung in die Funktionalanalysis} (Mannheim: BI).

\bibitem{joergens}
Joergens K  1964 {\em Spectral Theory of Second-Order Ordinary 
Differential Operators} Lectures delivered at Aarhus Universitet, 
(Mathematisk Institut Aarhus Universitet). 

\bibitem{kato} Kato T 1980 {\em Perturbation Theory for Linear 
Operators} (Berlin: Springer).

\bibitem{kojima0} Kojima Y 1992 {\em Equations governing the nonradial 
oscillations of a slowly rotating relativistic star} Phys. Rev. D, 46, 
4289-4303.

\bibitem{kojima} Kojima Y 1998 {\em Quasi-toroidal oscillations in rotating 
relativistic stars} 
Mon. Not. R. Astron. Soc., 293, 49-52.

\bibitem{Levinson} 
        Levinson N., 1948, 
        Duke Math. J., 15, 111. 

\bibitem{lockitch} Lockitch K H, Andersson N, Friedman J L 
2001 {\em Rotational modes of relativistic stars: Analytic results},
Phys. Rev. D, 63, 024019-1 - 024019-26.

\bibitem{lindblomowenmorsink} Lindblom L, Owen B J and Morsink S M 1998
{\em Gravitational radiation instability in hot young neutron stars}, 
Phys. Rev. Lett., 80, 4843-4846.

\bibitem{marden} Marden M 1989 
                 { \em Geometry of polynomials}
                 (Providence: AMS).

\bibitem{TOV}
        Misner C.W., Thorne K.S. and Wheeler J.A., 1973,
         {\em Gravitation} W.H.Freeman.


\bibitem{Owen98}
        Owen B, Lindblom L, Cutler C, Schutz B F, Vecchio A.
        and Andersson N 1998, {\em 
        Gravitational waves from hot young rapidly rotating neutron stars}
        Phys. Rev. D, 58, 084020-1 - 084020-15, gr-qc/9804044.

\bibitem{reedsimon} Reed M and Simon B 1980, 1975, 1979, 1978 {\em
 Methods of Mathematical Physics Volume I, II, III, IV} (New York:
 Academic).

\bibitem{riesznagy} 
        Riesz F and  Sz-Nagy B 1955 {\em Functional Analysis} 
        (New York: Unger). 

\bibitem{ruoffkokkotas} Ruoff J and Kokkotas K D 2001 {\em On the r-mode 
spectrum of relativistic stars in the low-frequency approximation},
M.N.R.A.S., 328, 678-688, gr-qc/0101105. 

\bibitem{ruoff} Ruoff J and Kokkotas K D 2002 {\em On the r-mode spectrum
of relativistic stars: Inclusion of the radiation reaction}, MNRAS in press,
gr-qc/0106073.

\bibitem{SV83}
        Schutz B F and Verdaguer E. 1983, 
{\em Normal modes of Bardeen discs - II. A sequence of n=2 
polytropes}, M.N.R.A.S., 202, 881-901.

\bibitem{weidmann} Weidmann J 1976 {\em Lineare Operatoren in
 Hilbertr\"{a}umen} (Teubner: Stuttgart).

\bibitem{yosida} Yosida K 1980 {\em Functional Analysis} (Berlin:
 Springer).

\end{thebibliography}
\end{document}